\documentclass[
    onecolumn,
	prd,
	amssymb,
	preprintnumbers,superscriptaddress,
	nofootinbib,notitlepage]{revtex4-2}
\pdfoutput=1
\usepackage{paralist}
\usepackage{graphicx}
\usepackage{enumitem}
\usepackage{latexsym}
\usepackage{amsfonts}
\usepackage{amssymb}
\usepackage{xcolor}
\usepackage[export]{adjustbox}
\usepackage{amsmath}
\usepackage[thinlines]{easytable}
\usepackage{slashed}
\usepackage{dcolumn}
\usepackage{verbatim}
\usepackage{float}
\usepackage{multirow}
\usepackage{xspace}
\usepackage[normalem]{ulem}
\usepackage[
pdfauthor={Jeremy Sakstein}]{hyperref}
\usepackage{tabularx}
\usepackage{lettrine}
\usepackage{setspace}
\usepackage{float}
\input Zallman.fd

\LettrineTextFont{\itshape}

\newcommand{\beq}{\begin{equation}}
\newcommand{\eeq}{\end{equation}}
\newcommand{\bea}{\begin{eqnarray}}
\newcommand{\eea}{\end{eqnarray}}

\newcommand{\refjnl}[1]{{\rm#1}}

\def\mnras{\refjnl{MNRAS}}             
\def\nat{\refjnl{Nature}}              

\usepackage{pdfbase}[2017/03/16]
\usepackage{xparse,ocgbase}
\usepackage{xcolor,calc}
\usepackage{tikzpagenodes,linegoal}
\usetikzlibrary{calc}
\usepackage{tcolorbox}

\ExplSyntaxOn
\let\tpPdfLink\pbs_pdflink:nn
\let\tpPdfAnnot\pbs_pdfannot:nnnn\let\tpPdfLastAnn\pbs_pdflastann:
\let\tpAppendToFields\pbs_appendtofields:n
\def\tpPdfXform{\pbs_pdfxform:nnnnn{1}{1}{}{}}
\let\tpPdfLastXform\pbs_pdflastxform:
\let\cListSet\clist_set:Nn\let\cListItem\clist_item:Nn
\ExplSyntaxOff

\usepackage{pdfbase}[2017/03/16]
\usepackage{xparse,ocgbase}
\usepackage{xcolor,calc}
\usepackage{tikzpagenodes,linegoal}
\usetikzlibrary{calc}
\usepackage{tcolorbox}

\ExplSyntaxOn
\let\tpPdfLink\pbs_pdflink:nn
\let\tpPdfAnnot\pbs_pdfannot:nnnn\let\tpPdfLastAnn\pbs_pdflastann:
\let\tpAppendToFields\pbs_appendtofields:n
\def\tpPdfXform{\pbs_pdfxform:nnnnn{1}{1}{}{}}
\let\tpPdfLastXform\pbs_pdflastxform:
\let\cListSet\clist_set:Nn\let\cListItem\clist_item:Nn
\ExplSyntaxOff

\makeatletter
\NewDocumentCommand{\tooltip}{%
  ssssO{\ifdefined\@linkcolor\@linkcolor\else blue\fi}mO{yellow!20}mO{0pt,0pt}%
}{{%
  \leavevmode%
  \IfBooleanT{#2}{%
    \ocgbase@new@ocg{tipOCG.\thetcnt}{%
      /Print<</PrintState/OFF>>/Export<</ExportState/OFF>>%
    }{false}%
    \xdef\tpTipOcg{\ocgbase@last@ocg}%
    \ocgbase@add@ocg@to@radiobtn@grp{tool@tips}{\ocgbase@last@ocg}%
  }%
  \tpPdfLink{%
    \IfBooleanTF{#4}{%
      /Subtype/Link/Border[0 0 0]/A <</S/SetOCGState/State [/Toggle \tpTipOcg]>>
    }{%
      /Subtype/Screen%
      /AA<<%
        \IfBooleanTF{#3}{%
          /E<</S/SetOCGState/State [/Toggle \tpTipOcg]>>%
        }{%
          \IfBooleanTF{#2}{%
            /E<</S/SetOCGState/State [/ON \tpTipOcg]>>%
            /X<</S/SetOCGState/State [/OFF \tpTipOcg]>>%
          }{
            \IfBooleanTF{#1}{%
              /E<</S/JavaScript/JS(%
                var fd=this.getField('tip.\thetcnt');%
                if(typeof(click\thetcnt)=='undefined'){%
                  var click\thetcnt=false;%
                  var fdor\thetcnt=fd.rect;var dragging\thetcnt=false;%
                }%
                if(fd.display==display.hidden){%
                  fd.delay=true;fd.display=display.visible;fd.delay=false;%
                }else{%
                  if(!click\thetcnt&&!dragging\thetcnt){fd.display=display.hidden;}%
                  if(!dragging\thetcnt){click\thetcnt=false;}%
                }%
                this.dirty=false;%
              )>>%
            }{%
              /E<</S/JavaScript/JS(%
                var fd=this.getField('tip.\thetcnt');%
                if(typeof(click\thetcnt)=='undefined'){%
                  var click\thetcnt=false;%
                  var fdor\thetcnt=fd.rect;var dragging\thetcnt=false;%
                }%
                if(fd.display==display.hidden){%
                  fd.delay=true;fd.display=display.visible;fd.delay=false;%
                }%
               this.dirty=false;%
              )>>%
              /X<</S/JavaScript/JS(%
                if(!click\thetcnt&&!dragging\thetcnt){fd.display=display.hidden;}%
                if(!dragging\thetcnt){click\thetcnt=false;}%
                this.dirty=false;%
              )>>%
            }%
            /U<</S/JavaScript/JS(click\thetcnt=true;this.dirty=false;)>>%
            /PC<</S/JavaScript/JS (%
              var fd=this.getField('tip.\thetcnt');%
              try{fd.rect=fdor\thetcnt;}catch(e){}%
              fd.display=display.hidden;this.dirty=false;%
            )>>%
            /PO<</S/JavaScript/JS(this.dirty=false;)>>%
          }%
        }%
      >>%
    }%
  }{{\color{#5}#6}}%
  \sbox\tiptext{%
    \IfBooleanT{#2}{%
      \ocgbase@oc@bdc{\tpTipOcg}\ocgbase@open@stack@push{\tpTipOcg}}%
    \tcbox[colframe=black,colback=#7,size=fbox,arc=1ex,sharp corners=southwest]{#8}%
    \IfBooleanT{#2}{\ocgbase@oc@emc\ocgbase@open@stack@pop\tpNull}%
  }%
  \cListSet\tpOffsets{#9}%
  \edef\twd{\the\wd\tiptext}%
  \edef\tht{\the\ht\tiptext}%
  \edef\tdp{\the\dp\tiptext}%
  \tipshift=0pt%
  \IfBooleanTF{#2}{%
    \setlength\whatsleft{\linegoal}%
  }{%
    \measureremainder{\whatsleft}%
  }%
  \ifdim\whatsleft<\dimexpr\twd+\cListItem\tpOffsets{1}\relax%
    \setlength\tipshift{\whatsleft-\twd-\cListItem\tpOffsets{1}}\fi%
  \IfBooleanF{#2}{\tpPdfXform{\tiptext}}%
  \raisebox{\heightof{#6}+\tdp+\cListItem\tpOffsets{2}}[0pt][0pt]{%
    \makebox[0pt][l]{\hspace{\dimexpr\tipshift+\cListItem\tpOffsets{1}\relax}%
    \IfBooleanTF{#2}{\usebox{\tiptext}}{%
      \tpPdfAnnot{\twd}{\tht}{\tdp}{%
        /Subtype/Widget/FT/Btn/T (tip.\thetcnt)%
        /AP<</N \tpPdfLastXform>>%
        /MK<</TP 1/I \tpPdfLastXform/IF<</S/A/FB true/A [0.0 0.0]>>>>%
        /Ff 65536/F 3%
        /AA <<%
          /U <<%
            /S/JavaScript/JS(%
              var fd=event.target;%
              var mX=this.mouseX;var mY=this.mouseY;%
              var drag=function(){%
                var nX=this.mouseX;var nY=this.mouseY;%
                var dX=nX-mX;var dY=nY-mY;%
                var fdr=fd.rect;%
                fdr[0]+=dX;fdr[1]+=dY;fdr[2]+=dX;fdr[3]+=dY;%
                fd.rect=fdr;mX=nX;mY=nY;%
              };%
              if(!dragging\thetcnt){%
                dragging\thetcnt=true;Int=app.setInterval("drag()",1);%
              }%
              else{app.clearInterval(Int);dragging\thetcnt=false;}%
              this.dirty=false;%
            )%
          >>%
        >>%
      }%
      \tpAppendToFields{\tpPdfLastAnn}%
    }%
  }}%
  \stepcounter{tcnt}%
}}
\makeatother
\newsavebox\tiptext\newcounter{tcnt}
\newlength{\whatsleft}\newlength{\tipshift}
\newcommand{\measureremainder}[1]{%
  \begin{tikzpicture}[overlay,remember picture]
    \path let \p0 = (0,0), \p1 = (current page.east) in
      [/utils/exec={\pgfmathsetlength#1{\x1-\x0}\global#1=#1}];
  \end{tikzpicture}%
}

\newcommand{\dd}{\mathrm{d}}

\DeclareRobustCommand{\okina}{%
  \raisebox{\dimexpr\fontcharht\font`A-\height}{%
    \scalebox{0.8}{`}%
  }%
}

\allowdisplaybreaks

\begin{document}

\title{Neutron Stars in Aether Scalar-Tensor Theory}

\author{Christopher Reyes}
    \email{cmreyes3@hawaii.edu}
    \affiliation{Department of Physics \& Astronomy, University of Hawai\okina i, Watanabe Hall, 2505 Correa Road, Honolulu, HI, 96822, USA}
    \author{Jeremy Sakstein} \email{sakstein@hawaii.edu}
\affiliation{Department of Physics \& Astronomy, University of Hawai\okina i, Watanabe Hall, 2505 Correa Road, Honolulu, HI, 96822, USA}

\date{\today}

\begin{abstract}
Aether Scalar-Tensor theory is a modification of general relativity proposed to explain galactic and cosmological mass discrepancies conventionally attributed to dark matter.~The theory is able to fit the cosmic microwave background and the linear matter power spectrum.~In this work, we derive the Tolman-Oppenheimer-Volkoff equation in this theory and solve it for realistic nuclear equations of state to predict the mass-radius relation of neutron stars.~We find solutions that are compatible with all current observations of neutron stars.
\end{abstract}

\maketitle

\section{Introduction}
\label{sec:intro}

Mass discrepancies are abundant in the Universe from cosmological to galactic scales.~General relativity (GR) applied solely to the luminous matter of astronomical systems on these scales fails to reproduce their observed properties.~Examples  include the cosmic microwave background (CMB), the large scale structure (LSS) of the Universe, and the flattening of galactic rotation curves (see \cite{Famaey:2011kh} for a comprehensive discussion).~There are two hypotheses for explaining the mass discrepancies in these systems.~The first is that there is an additional component of non-luminous dark matter (DM) that contributes to the gravitational potential of these systems but not their brightness.~The second is that Newton's law of gravitation is modified.~There is also the possibility that both of these phenomena are present in some capacity.~

The dark matter paradigm is resoundingly successful at explaining observations on large scales where cosmological perturbations are linear such as the CMB and LSS.~On smaller, non-linear scales, it is less successful.~Specifically, it has difficulty in explaining the observed cored density profiles of some galaxies \cite{Ostriker:2003qj,DelPopolo:2021bom}, the masses and phase-space distribution of Milky Way satellite galaxies \cite{2011MNRAS.415L..40B,2012MNRAS.423.1109P,Pawlowski:2013cae,2014Natur.511..563I,Ibata:2013rh}, and the  dynamics of tidal dwarf galaxies \cite{Gentile:2007gp,Kroupa:2012qj,Kroupa:2014ria} without invoking highly-uncertain baryonic physics.~In addition, DM is unable to predict the empirical Tully-Fisher \cite{Tully:1977fu} and Faber-Jackson \cite{Faber:1976sn} relations from first principles.~In contrast, the modified gravity paradigm is able to fit our observations on scales smaller than galaxy clusters \cite{Famaey:2011kh}, but has difficulty fitting observations on larger scales, with the CMB and LSS being notable examples.

The quintessential paradigm for modified gravity resolutions of the mass-discrepancy problem is modified Newtonian dynamics (MOND) \cite{Milgrom:1983ca,Bekenstein:1984tv,Milgrom:2009ee,Famaey:2011kh,Banik:2021woo}.~In this framework, the Poisson equation for determining the gravitational potential sourced by matter distributions is modified such that Newton's law of gravitation is altered in systems with accelerations smaller than the scale $a_0\sim 1.2\times10^{-10}$ m/s$^2$.~MOND is successful at explaining the small scale phenomenon discussed above, but it is a non-relativistic theory and lacks the fundamental relativistic completion to make predictions for gravitational lensing and  cosmological observables.~Several relativistic completions have been proposed \cite{Bekenstein:1988zy,Sanders:1996wk,Bekenstein:2004ne,Moffat:2005si,Navarro:2005ux,Zlosnik:2006zu,Sanders:2005vd,Milgrom:2009gv,Babichev:2011kq,Deffayet:2011sk,Blanchet:2011wv,Sanders:2011wa,Mendoza:2012hu,Woodard:2014wia,Khoury:2014tka,Blanchet:2015sra,Hossenfelder:2017eoh,Burrage:2018zuj,Milgrom:2019rtd,DAmbrosio:2020nev,Kading:2023hdb}, but many of them are unable to successfully predict the observed CMB or linear matter power spectra.~In addition, several predict that the speed of gravitational waves differs appreciably from the speed of light, a scenario which is now excluded by GW170817 --- the near-coincident observation of gravitational waves and light from a binary neutron star merger that constrains any difference between their speeds to be smaller than one part in $10^{16}$ \cite{LIGOScientific:2017vwq,Sakstein:2017xjx,Baker:2017hug,Creminelli:2017sry,Ezquiaga:2017ekz}.

Recently, Skordis and Z\l{}o\'{s}nik have developed Aether Scalar-Tensor theory (AeST) \cite{Skordis:2020eui}.~AeST is a Scalar-Vector-Tensor theory that reproduces the MOND force law in the weak-field quasi-static limit, and provides a good fit to the CMB and linear matter power spectra.~The theory propagates six degrees of freedom \cite{Bataki:2023uuy}, all of which can be healthy given suitable parameter choices.~The spin-2 degrees of freedom propagate luminally.~The predictions of the theory have been studied in cosmological, Newtonian, and black hole settings \cite{Mistele:2021qvz,Kashfi:2022dyb,Bernardo:2022acn,Mistele:2023paq,Mistele:2023fwd,Verwayen:2023sds,Durakovic:2023out,Llinares:2023lky,Tian:2023gjt,Rosa:2023qun,Rosa:2024fwc}.

In this work, we study neutron stars (NSs) in AeST.~We derive the Tolman-Oppenheimer-Volkoff (TOV) equations resulting from the theory, and solve them numerically using realistic nuclear equations of state to find the mass-radius relations.~We find that these deviate from the predictions of GR but can be made consistent with all current  observations of neutron stars for a large range of suitable values of the free parameters.~Our study provides further evidence in support of AeST as a viable theory of gravity.

This paper is organized as follows.~In section \ref{sec:Aest_review} we review the salient features of AeST.~In section~\ref{sec:TOV} we derive the TOV equations under AeST and give our numerical integration procedure.~In section~\ref{sec:results} we solve the TOV equations using realistic equations of state.~We discuss our results and conclude in section~\ref{sec:conclusions}.

\section{Aether Scalar-Tensor Theory}
\label{sec:Aest_review}

Aether-Scalar tensor theory is a tensor-vector-scalar theory of the space-time metric $g_{\mu\nu}$, a unit-timelike vector $A_\mu$ called the aether, and a scalar $\phi$.~The action is \cite{Skordis:2020eui}
\begin{align}
S&=\int\mathrm{d}^4x\frac{\sqrt{-g}}{16\pi G}\left[R-\frac{K_B}{2}{F}^{\mu\nu}{F}_{\mu\nu}-\lambda(A_\mu A^\mu +1)+(2-K_B)(2J^\mu\nabla_\mu\phi-\mathcal{Y})-\mathcal{F}(\mathcal{Y},\mathcal{Q})\vphantom{\frac{K_B}{2}{F}^{\mu\nu}{F}_{\mu\nu}}\right] +S_{\rm m}[g],\label{eq:action}
\end{align}
where $F_{\mu\nu}=2\nabla_{[\mu}A_{\nu]}$, $\nabla_\mu$ is the connection compatible with $g_{\mu\nu}$, $\lambda$ is a Lagrange multiplier that enforces the unit-timelike constraint for the vector $A_\mu$, $J_\mu=A^\nu\nabla_\nu A_\mu$, $\mathcal{Q}=A^\mu\nabla_\mu\phi$, $\mathcal{Y}=q^{\mu\nu}\nabla_\mu\phi\nabla_\nu\phi$ with $q_{\mu\nu}=g_{\mu\nu}+A_\mu A_\nu$.~The function $\mathcal{F}(\mathcal{Y},\mathcal{Q})$ is arbitrary and $K_B$ is a constant.~The free function $\mathcal{F}(\mathcal{Y},\mathcal{Q})$ determines the local behavior of gravity, and is used to reproduce the MOND behavior.~In this work we will study neutron stars, where the typical acceleration is much greater than the MOND scale $a_0=1.2\times10^{-10}$ m/s$^2$.~In this regime $\mathcal{F}(\mathcal{Y},\mathcal{Q})=(2-K_{B})\lambda_{\rm s}\mathcal{Y}$ with $\lambda_{\rm s}$ a free parameter that must be positive in order for the theory to be stable on Minkowski space \cite{Skordis:2020eui,Skordis:2021mry,Verwayen:2023sds}.~Newton's constant as measured by solar system experiments is given by $G_N=(1+\lambda_{\rm s}^{-1})(1-K_B/2)^{-1}G$ \cite{Skordis:2020eui,Skordis:2021mry}.~This mandates that $0<K_B<2$ so that the graviton and vector have correct-sign kinetic terms.~An additional term in $\mathcal{F}(\mathcal{Y},\mathcal{Q})$ of the form $ \mathcal{K}_2(\mathcal{Q}-\mathcal{Q}_0)^2$ is necessary to achieve consistency with cosmological observations \cite{Skordis:2020eui}.~This induces a mass for the metric potentials $\mu$ that must be taken to satisfy $\mu\sim$ Mpc$^{-1}$ to reproduce observations \cite{Skordis:2020eui}.~This term is irrelevant for neutron stars because $\mu{R}_{\rm NS}\ll1$  with $R_{\rm NS}\sim10$ km a typical NS radius so we neglect it in what follows.

The field equations can be obtained by varying the action with respect the metric, the scalar and, the vector.~The metric variation gives us the modified Einstein equations:
\begin{eqnarray}
    &&G_{\mu\nu}  =8\pi GT^{\rm AeST}_{\mu\nu}+ 8\pi G T_{\mu\nu}\label{eq:Modified},
\end{eqnarray}
where $T_{\mu\nu}$ is the  matter energy-momentum tensor and $T^{\rm AeST}_{\mu\nu}$ is the effective AeST energy momentum tensor given by: 
\begin{align}
 8\pi G T^{\rm AeST}_{\mu\nu} &=   K_BF_{\mu}^\alpha F_{\nu\alpha} - 
    (2-K_{B}) \{ 2J_{(\mu}\nabla_{\nu)}\phi - A_{\mu}A_{\nu}\nabla^{\alpha}\nabla_{\alpha}\phi \nonumber 
     + 2\left[A_{(\mu}\nabla_{\nu)}A_{\alpha} - A_{(\mu}\nabla_{|\alpha|}A_{\nu)}\right]\nabla^{\alpha}\phi\}   \nonumber \\ 
     &+ \mathcal{F}_{\mathcal{Q}}A_{(\mu}\nabla_{\nu)}\phi + (2-K_{B}+ \mathcal{F}_{\mathcal{Y}}) [ \nabla_{\mu}\phi\nabla_{\nu}\phi + 2\mathcal{Q}A_{(\mu}\nabla_{\nu)}\phi]  
      +\lambda A_{\mu}A_{\nu}\nonumber\\
      &-  \frac12g_{\mu\nu}\left(\frac{K_B}{2}{F}^{\mu\nu}{F}_{\mu\nu}+\lambda(A_\mu A^\mu +1)-(2-K_B)(2J^\mu\nabla_\mu\phi-\mathcal{Y})+\mathcal{F}(\mathcal{Y},\mathcal{Q})\vphantom{\frac{K_B}{2}{F}^{\mu\nu}{F}_{\mu\nu}}\right),
 \end{align}   
where $\mathcal{F}_{\mathcal{Y}}= \partial _{\mathcal{Y}}\mathcal{F}$,  $\mathcal{F}_{\mathcal{Q}}= \partial _{\mathcal{Q}}\mathcal{F}$.~Varying the action with respect to the scalar field we obtain 
\begin{align}
\nabla_{\mu}\mathcal{J}^{\mu} =0,\label{eq:scalar}
\end{align}
where $\mathcal{J}^{\mu}$ = $(2-K_{B})J^{\mu} -  (2-K_{B}+\mathcal{F}_{\mathcal{Y}})q^{\alpha\mu}\nabla_{\alpha}\phi - \mathcal{F}_{\mathcal{Q}}A^{\mu}/2$.~The vector variation yields 
\begin{eqnarray}
&&K_{B}\nabla_{\nu}F^{\nu\mu} + (2-K_{B})\left[(\nabla^{\nu}A^{\mu})\nabla_{\nu}\phi - \nabla_{\nu}(A^{\nu}\nabla^{\mu}\phi) \right] 
 -\left[ (2-K_{B} + \mathcal{F}_{\mathcal{Y}}\mathcal{Q}) + \mathcal{F}_{\mathcal{Q}}/2\right]\nabla^{\mu}\phi - \lambda A^{\mu} = 0\label{eq:vector}.
\end{eqnarray}
and, lastly, varying the action with respect the Lagrange multiplier $\lambda$ yields the constraint
\begin{align}
    A^{\mu}A_{\mu} +1 =0.\label{eq:constrain}
\end{align}

\section{Modified Tolman-Oppenheimer-Volkoff Equations}
\label{sec:TOV}

We will consider a static spherically symmetric NS, so we can write the metric as 
\begin{equation}
    \dd s^{2}= -e^{\alpha(r)}\dd t^{2}
     +e^{2\gamma(r)}\dd r^{2} +r^{2}
 \dd\theta^{2} + r^2\sin^{2}\theta\dd\phi^2
 \label{eq:metric}
\end{equation}
where, in accordance with the symmetries, $\alpha$ and $\gamma$ are the metric components and are functions of $r$ only.~We similarly write the scalar as $\phi$ = $\phi(r)$.~To satisfy the constraint in Eq.~\eqref{eq:constrain} we write the vector field as
\begin{equation}
    A_{\mu}dx^{\mu}= e^{\frac{\alpha(r)}{2}}\dd t,  \label{eq:vectorAnt}
\end{equation}
We treat the matter as a perfect fluid with energy-momentum tensor 
\begin{equation}
    T_{\mu\nu} = \left( \rho +P  \right)u_{\mu}u_{\nu} 
    + Pg_{\mu\nu},\label{eq:EM}
\end{equation}
where $\rho$ and P are the energy density and pressure of the fluid with four-velocity
\begin{equation}
u^{\mu}\partial_{\mu} = e^{-\frac{\alpha(r)}{2}}\partial_{t}.~\label{eq:vel}
\end{equation}
Finally, using the conservation of the energy-momentum tensor of the fluid, $\nabla^{\mu}T_{\mu\nu} = 0$, which follows from taking a covariant derivative of Eq.~\eqref{eq:Modified}, imposing the Bianchi identity, and eliminating $\phi$, $A^\mu$, and $\lambda$  using their equations of motion, we obtain the continuity equation:
\begin{align}
    \frac{dP}{dr} = - \frac12[P(r)+\rho(r)]\alpha^{\prime}(r),\label{eq:dp}
\end{align}
where a prime denotes a derivative with respect to $r$.

Combining the ansatzes for the metric, vector, and scalar above with Eqs.~\eqref{eq:Modified},~\eqref{eq:scalar}, and~\eqref{eq:vector}, we obtain differential equations for the metric components $\alpha(r)$ and  $\gamma(r)$.~These are long and cumbersome so we give their expressions and subsequent manipulations in Appendix~\ref{sec:Appendix}.~We provide a code that performs the derivation in full at the following URL:~\href{https://zenodo.org/records/12537564}{https://zenodo.org/records/12537564}.~Here, we describe the salient features.~The vector equation yields an expression for the Lagrange multiplier $\lambda$, which can be used to eliminate it from all other equations.~The scalar equation yields $\phi(r)\propto \alpha(r)$ up to an irrelevant constant, and, upon using this in the $rr$-component of the modified Einstein equations, one obtains an algebraic relation for $\gamma(r)$ as a function of $\alpha(r)$ and $P(r)$.~All of these can be used in the $tt$-component of the modified Einstein equations to find a second-order differential equation for $\alpha(r)$ given in equation~\eqref{eq:alpOD}.~The system is closed once a barotropic equation of state $P(\rho)$, discussed in the next section, is specified and substituted into the continuity equation.~The resulting differential equations for $\alpha(r)$ and $\rho(r)$ must be solved numerically to determine the structure of the star.~The resulting solutions for $\alpha(r)$ and $\rho(r)$ fully determine $\phi(r)$, $\gamma(r)$, and $P(r)$.

The details of our integration are as follows.~First, for numerical stability we expand the equations to $\mathcal{O}(r^2)$ to find 
\begin{align}
    \alpha(r) &= \alpha_{c} + \frac{8\pi G_N(3p_{c}+\rho_{c})}{3}r^{2}
    + \mathcal{O}(r^{4})\label{eq:alpcen}\\
    \rho(r) &= \rho_{c} - \frac{4\pi G_N(p_{c}+\rho_{c})(3p_{c}+\rho_{c})}{3\,{\dd{P}}/{\dd\rho}|_c}r^{2}
    + \mathcal{O}\left(r^{4}\right),\label{eq:rhocen}
\end{align}
where a subscript $c$ indicates central values and we imposed the conditions $P(0)=P_c$ and $\rho'(0)=0$ corresponding to spherical symmetry.~The derivation of these expressions is given in our reproduction package~\cite{reyes_2024_12537564}.~Stable stellar configurations can only exist when $\rho'(r)<0$ so that the pressure and density decrease away from the center.~This condition is satisfied for all choices of $K_B$ and $\lambda_s$ when $\lambda_s>0$.

To construct NS models we integrate the system of equations from a small value $r_i=0.1$ cm near the center of the star with $\alpha(r_i)$ and $\rho(r_i)$ given by equations~\eqref{eq:alpcen} and \eqref{eq:rhocen} up to the radius of the NS, $r=R$, defined as $P(R) = 0$.~Next, using the values of $\alpha(R)$ and $\alpha^{\prime}(R)$ we solve the system of equations outside of the star by setting $P(r)$ and $\rho(r)$ to zero.~To determine the neutron star's mass $M$ we match our numerical solution for the metric potential $\alpha(r)$ to its asymptotic expansion
\begin{align}
\label{eq:alpa_asymptotic}
    \lim_{r\to\infty}{\alpha(r)} = A - \frac{{2}G_NM}{r} -\frac{1}{2}\left(\frac{{2}G_NM}{r}\right)^{2} + \mathcal{O}\left(\frac{1}{r^{3}}\right),
\end{align}
which we derive in Appendix~\ref{sec:Appendix} and \cite{reyes_2024_12537564}.~Here, $A$ is an integration constant that arises because the value of $\alpha_c$ is arbitrary, with different values corresponding to constant rescalings of the time coordinate.~The asymptotic expansion for $g_{rr}$ has the same form as GR to $\mathcal{O}(r^{-1})$, as can be seen by inserting Eq.~\eqref{eq:alpa_asymptotic} into Eq.~\eqref{eq:Grr}.~Strictly speaking, the \textit{asymptotic} expansion is really an expansion in an intermediate regime valid at radii larger than the NS radius ($\sim 10$ km) but smaller than the radius where the theory transitions to the MOND regime i.e., the radius where $\mathcal{F}(\mathcal{Y},\mathcal{Q})$ becomes non-linear in $\mathcal{Y}$ ($\sim $ kpc).~This distinction is unimportant for determining the NS mass since the metric rapidly approaches post-Minkowskian limit after a few Schwarzschild radii ($\sim6$ km).

 \section{Results}
 \label{sec:results}

We solved the TOV equations following the procedure above assuming three commonly used barotropic equations of state --- APR \cite{PhysRevC.58.1804}, SLY4 \cite{Chabanat:1997un}, and MP1 \cite{Lattimer:2000nx,Lattimer:2006xb}~The first two are now excluded in GR by observations of GW170817 \cite{Radice:2017lry} but not in AeST because this event has not yet been modeled in this theory.~We do not expect qualitatively different results with other realistic equations of state.~The resultant mass-radius relations for varying values of the free parameters $K_B$ and $\lambda_{\rm s}$ are shown in figure~\ref{fig:MR_APR} for the APR EOS.~The relations for SLY4 and MP1 are visually similar so are shown in Appendix~\ref{app:MR}.~Also shown are X-ray measurements of the masses and radii of confirmed neutron stars.~These are:~PSR J0740+6620 \cite{Miller_2021}, PSR J0030+0451 \cite{Riley_2019}, and PSR J0348+0432 \cite{Antoniadis_2013}.~GR is compatible with all of these observations for all three equations of state.~The figures demonstrate that there are parameters where AeST is similarly compatible.~The parameter $\lambda_{s}$ is necessary for the existence of a \textit{tracking mechanism} that suppresses post-Newtonian AeST effects in the solar system, with $\lambda_{s}\gg1$ necessary to satisfy experimental bounds \cite{Skordis:2020eui}.~The large values of $\lambda_{s}$ needed to accommodate neutron star observations are then reasonable in the context of this theory.~Similarly, $K_B\sim0.1$ provides an excellent fit to the CMB and matter power spectra, so the values studied in the figures are reasonable.~In all cases, the maximum neutron star mass is lighter than predicted by GR, consistent with the expectation that the additional AeST degrees of freedom enhance the force of gravity. 

\begin{figure}
    \centering
\includegraphics[width=0.9\textwidth]{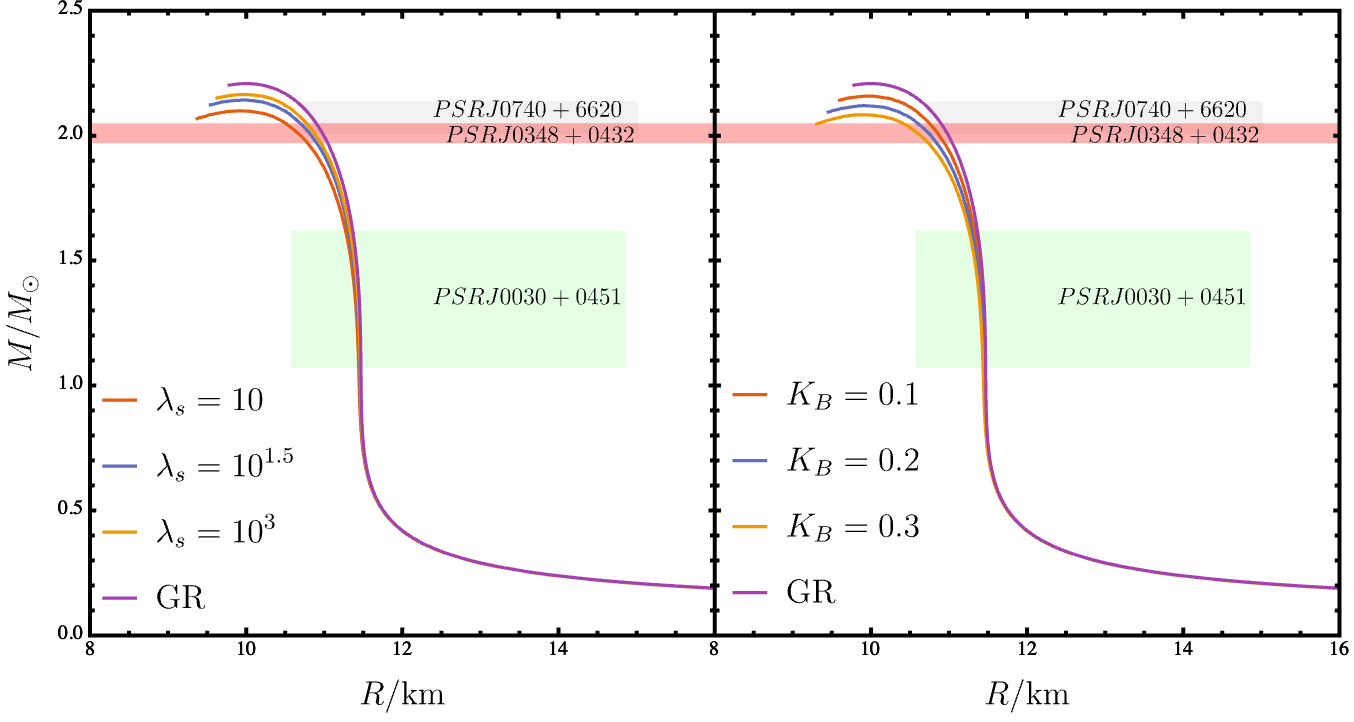}
    \caption{Mass-radius relations for the AeST parameters indicated in the figure assuming the APR EOS.~The left panel shows the effect of varying $\lambda_{s}$ at fixed $K_{B}=0.1$, and the right panel shows the effect of varying $K_{B}$ at fixed $\lambda_{s}=10^2$.}
    \label{fig:MR_APR}
\end{figure}

\section{Discussion and Conclusions}
\label{sec:conclusions}

The development of Aether Scalar-Tensor theory was a major milestone for modified gravity as an explanation for mass discrepancies across the universe. The theory overcame several limitations of previous models, most notably it is able to fit the CMB and predicts gravitational waves that propagate luminary on cosmological backgrounds.~Our work has provided further evidence that AeST is a viable theory of gravity by demonstrating that neutron stars are both predicted to exist, and have masses and radii compatible with observations for a large range of reasonable model parameters.~Unfortunately, the equation of state for nuclear matter is presently unknown \cite{Lattimer:2012nd,Burgio:2021vgk,Lattimer:2021emm}, so it is not possible to bound the model parameters using masses and radii of individual objects. The natural next step is then to make predictions for quantities that are observable.~One possibility is to extend our work to derive the equations for slowly rotating neutron stars in AeST.~This would enable computations of \textit{universal relations} such as the I-Love-Q and $I$--$\mathcal{C}$ relations, which hold independently of the EOS and therefore remove its associated uncertainty \cite{Yagi:2013bca,Yagi:2013awa,Yagi:2016bkt,Doneva:2017jop}.~The formalism we have developed lays the foundation for such studies.~In addition, it would be interesting to derive the behavior of binary pulsars in AeST.~These systems have previously yielded strong bounds on other theories that include a constrained vector degree of freedom e.g., Einstein-Aether theory \cite{Yagi:2013ava,Yagi:2013qpa,Gupta:2021vdj} and TeVeS \cite{Kramer:2021jcw}.~Finally, we note that it may be possible to find other branches of neutron star solutions by generalizing the ansatzes we have made here, either by allowing a radial component for the vector in \eqref{eq:vectorAnt} or by endowing the scalar with time-dependence so that $\phi(t,r)=qt+\varphi(r)$.

\section*{Software}

Mathematica version 12.3.1.0, xCoba version 0.8.6,  xPerm version 1.2.3, xPert 1.0.6 , xTensor version 1.2.0, xTras version 1.4.2.

\section*{Acknowledgements}

We are grateful for discussions with Emanuele Berti, Constantinos Skordis, and Hector O.~C.~Silva.

\appendix

 \section{Derivation of the Tolman-Oppenheimer-Volkoff Equations}
 \label{sec:Appendix}

In this appendix we evaluate the equations for the metric~\eqref{eq:Modified}, vector~\eqref{eq:scalar}, and scalar~\eqref{eq:vector} given the ansatzes in section~\ref{sec:TOV} to derive some expressions given there.~A complete derivation of all formulae presented here can be found in our supplemental code release \cite{reyes_2024_12537564}.~

We begin with the scalar equation of motion \eqref{eq:scalar}, which yields
\begin{align}
    \label{eq:scalar_expression}
    \varphi '(r) = \frac{ \alpha '(r)}{2 (1+\lambda_{s})}, 
\end{align}
which was found by setting $\mathcal{J}^r=0$.~The more general solution where $\mathcal{J}^r$ is equal to a non-zero constant gives rise to a radial flux for the scalar's energy-momentum tensor at infinity, and is therefore unphysical \cite{Babichev:2015rva}.~Equation \eqref{eq:scalar_expression} implies that $\phi(r)\propto\alpha(r)$ up to an irrelevant integration constant.~Using this relation in the vector equation yields an expression for the Lagrange multiplier $\lambda$: 

\begin{align}
\label{eq:lambda_expression}
\lambda=\frac{e^{-2 \gamma (r)} \left(\alpha'(r) \left((K_{B}-2) r \alpha'(r)+(\lambda_{s} +1) K_{B} \left(2-r \gamma'(r)\right)\right)+(\lambda_{s} +1) K_{B} r \alpha''(r)\right)}{2 (\lambda_{s} +1) r}
\end{align}
Equations \eqref{eq:scalar_expression} and \eqref{eq:lambda_expression} can be used to simplify the modified Einstein equations \eqref{eq:Modified} to find the $tt$-component: 

\begin{align}
&8 \lambda_{s} +4 \lambda_{s}  K_{B} r^2 \alpha ''(r)+r^2 (\lambda_{s}  K_{B}+2) \alpha '(r)^2-4 r (\lambda_{s}  K_{B}+2) \alpha '(r) \left(r \gamma '(r)-2\right)+8 r^2 \alpha ''(r)\nonumber\\&-16 (\lambda_{s} +1) r \gamma '(r)-8 \lambda_{s}  e^{2 \gamma (r)}-8 e^{2 \gamma (r)}+8=-64 \pi  G_{N}(1-K_{B}/2)\lambda_{\rm s}  r^2 e^{2 \gamma (r)} \rho \label{eq:Gtt}
\end{align}
the $rr$-component: 
\begin{align}
   r \alpha^{\prime}(r)+1- e^{2\gamma(r) } +\frac{r^{2} (2+K_{B}\lambda_{s}) \alpha^{\prime2}(r) }{8 (1+\lambda_{s})} =  8 \pi  G_{N}(1-K_{B}/2)\left(1+\frac{1}{\lambda_{\rm s}}\right)^{-1} r^{2} P e^{2 \gamma(r)},\label{eq:Grr}
\end{align}
and the $\theta\theta$-component: 
\begin{align} 
  \frac{\lambda_{s}  (K_{B}-2)\lambda_{s} \alpha '(r)^2}{1+\lambda_{s}}+4 \alpha ''(r)+\alpha '(r) \left(\frac{4}{r}-4 \gamma '(r)\right)=64 \pi  G_{N}(1-K_{B}/2)\left(1+\frac{1}{\lambda_{\rm s}}\right)^{-1}  P e^{2 \gamma (r)}+\frac{8 \gamma '(r)}{r}.\label{eq:Gtheta}
\end{align}
All other components are either zero or equivalent to these.~The $\theta\theta$-equation does not yield an independent equation due to the isotropy of the fluid.~This can be verified explicitly by eliminating $\gamma(r)$ and $\gamma^{\prime}(r)$ and using the conservation of energy-momentum, after which it is found to be equivalent to the $tt$-equation.

Equation \eqref{eq:Grr} is algebraic in the metric potential $\gamma(r)$, so we can use it to eliminate $\gamma(r)$ in equation $\eqref{eq:Gtt}$ to find a second-order differential equation for $\alpha(r)$:
\begin{equation}
    \alpha ''(r)+\frac{\Upsilon(r;K_B,\lambda_{s})}{8 (\lambda_{\rm s} +1) r \left(\lambda_{\rm s} +1-4 \pi  G_{N} \lambda_{\rm s}  (K_B-2) r^2 P\right)}=0\label{eq:alpOD}
\end{equation}
with 
\begin{align}
\Upsilon(r;K_{B},\lambda_{s})&=-4 \pi  G_{N} (\lambda_{\rm s} +1) r \rho \left(r \alpha '(r)+2\right) \left(8 (\lambda_{\rm s} +1)+r \alpha '(r) \left(8 (\lambda_{\rm s} +1)+r (\lambda_{\rm s}  K_{B}+2) \alpha '(r)\right)\right)\\&+(\lambda_{\rm s} +1) \alpha '(r) \left(16 (\lambda_{\rm s} +1)+r \alpha '(r) \left(8 (\lambda_{\rm s} +1)+r (\lambda_{\rm s}  K_{B}+2) \alpha '(r)\right)\right)
    -4 \pi  G_{N} r P \left(48 (\lambda_{\rm s} +1)^2 \right)\nonumber\\&-4 \pi  G_{N} r P\left(r \alpha '(r) \left(24 (\lambda_{\rm s} +1) (\lambda_{\rm s} +\lambda_{\rm s}  K_{B}+3)+r \alpha '(r) \left(2 (\lambda_{\rm s} +1) (\lambda_{\rm s}  (11 K_{B}-4)+18)\right)\right)\right)\nonumber\\&
    -4 \pi  G_{N} r Pr^{2}\alpha^{\prime2}(r)\left(\left(\left(r (\lambda  K_{B}+2) (\lambda_{\rm s}  (2 K_{B}-1)+3) \alpha '(r)\right)\right)\right)\nonumber
\end{align}
This equation and the continuity equation for the fluid \eqref{eq:dp} are all that need to be solved to fully determine the structure of the star and the spacetime.~Once these are solved for $\alpha(r)$ and $P(r)$, the metric potential $\gamma(r)$ is determined via equation~\eqref{eq:Grr} and the scalar is determined by \eqref{eq:scalar_expression}.~

The one remaining item to address is the matching conditions at the NS surface, which may be non-trivial in modified gravity theories \cite{Rosa:2023tph}.~Examining the equations of motion, it is clear that the correct conditions are that all fields are continuous and smooth.~The ansatzes we have made --- standard TOV gauge for the metric, vector aligned with the time direction, all fields depending solely on the radial coordinate, and a perfect fluid energy-momentum tensor --- are valid throughout the entire spacetime, including at the surface of the star where the energy-momentum tensor and its covariant derivative falls to zero continuously.~This continuity ensures that there are no discontinuous quantities in the field equations, which then mandates that all other fields and their derivatives be continuous.~As mentioned in the main text, our numerical integration procedure ensures that $\alpha$, $P$, and $\rho$ are continuous and smooth.~The continuity and smoothness of $\phi$, $\lambda$, $\gamma$, and $A^\mu$ are ensured by this continuity via equations~\eqref{eq:scalar_expression},~\eqref{eq:lambda_expression},~\eqref{eq:Grr}, and~\eqref{eq:vectorAnt}.

As discussed in section~\ref{sec:TOV}, it is necessary to determine an asymptotic expansion for $\alpha(r)$ as $r\to\infty$.~This can be accomplished as follows.~Outside of the NS where $P$ and $\rho$ are zero, equation \eqref{eq:alpOD} simplifies to: 
\begin{align}
    2r\alpha^{\prime}(r) +r^{2}\alpha^{\prime2}(r) + \xi r^{3}\alpha^{\prime3}(r)
+r^{2}\alpha^{\prime\prime}(r)=0
\end{align}
where $\xi= \frac{2+K_{B}\lambda_{s}}{8(1+\lambda_{s})} $ making the substitution $Y=r\alpha(r)^{\prime}$ we obtain 
\begin{align}
    r\frac{dY}{dr} +  Y(1+ Y + \xi Y^{2})  &= 0\label{eq:dydr}\\
   \frac{d\alpha}{dY}+  \frac{1}{(1+ Y + \xi Y^{2})}  &= 0\label{eq:dady}
\end{align}
Following a similar procedure to reference \cite{Eling:2007xh} equations \eqref{eq:dydr} and \eqref{eq:dady} can be solved analytically in term of the roots of $(1+ Y + \xi Y^{2})$:
\begin{align}
   e^{\alpha} = e^{A}\left (\frac{1-Y/Y_{-}}{1-Y/Y_{+}}\right)^{\frac{-Y_{+}}{2+Y_{+}}}
\end{align}
and 
\begin{align}
   \frac{r_{\rm min}}{r} =\left( \frac{Y}{Y-Y_{-}}\right) \left (\frac{Y-Y_{-}}{Y-Y_{+}}\right)^{\frac{1}{2+Y_{+}}}
\end{align}
where $r_{\rm min}$ and A are integration constants and $Y_{\pm}$ = $(-1 \pm \sqrt{1-4\xi})/(2\xi)$. Expanding these equations at spatial infinity one finds
\begin{align}
   e^{\alpha} &= e^{A}\left(1 - \frac{{2}G_NM}{r} - \frac{\xi}{6}\left(\frac{{2}G_NM }{r}\right) ^{3}+ \dots\right) \\
   Y &=  \frac{{2}G_NM}{r} + \left(\frac{{2}G_NM }{r}\right) ^{2} +\left(1+\frac{\xi}{2}\right)\left(\frac{{2}G_NM }{r}\right) ^{3} +\dots
\end{align}
where is M is the mass of the star, which is related to $r_{\rm min}$ by
\begin{align}
   r_{\rm min} = \left(-Y_{+} \right)^{-1}\left(-1-Y_{+}\right)^{(1+Y_{+})/(2+Y_{+})}{2}G_NM
\end{align}
To first order this solution agrees with GR.

\section{Mass-Radius Relations for the APR and MP1 Equations of State}
\label{app:MR}

\begin{figure}
    \centering
\includegraphics[width=0.95\textwidth]{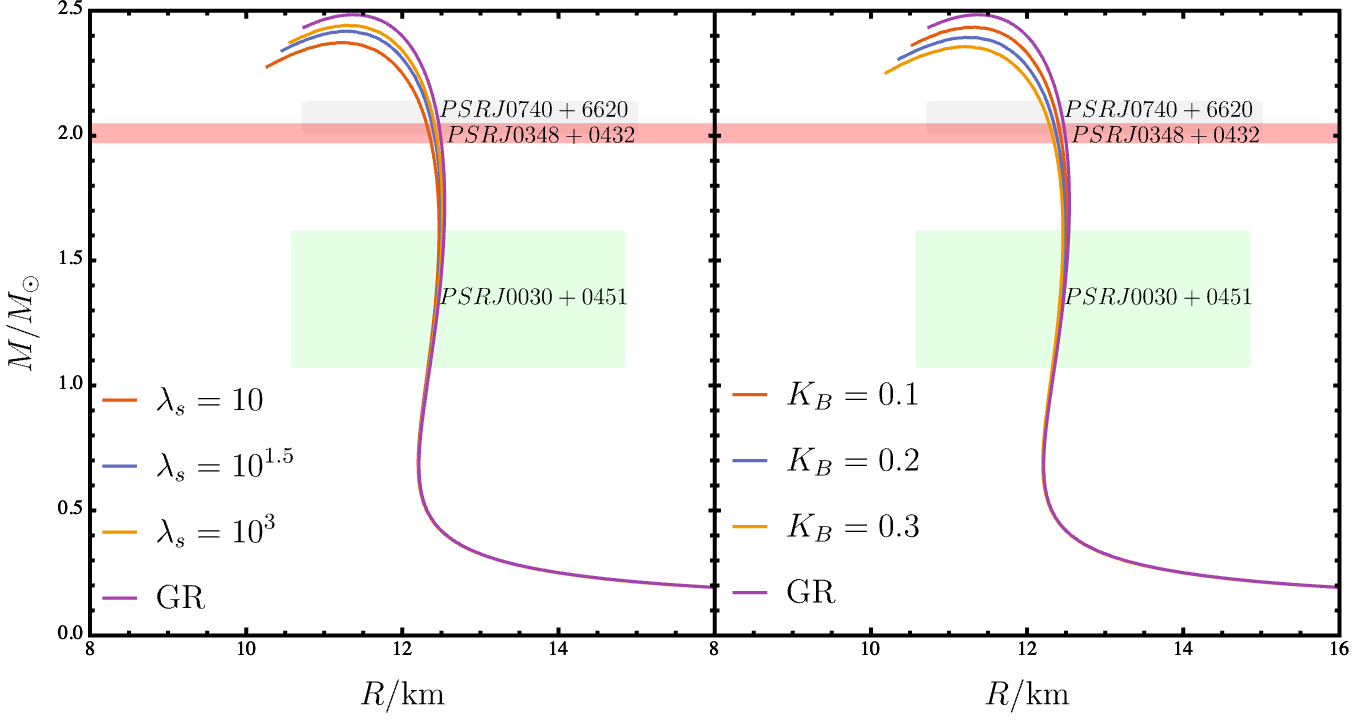}
    \caption{Same as figure~\ref{fig:MR_APR} but using the MP1 EOS.}
    \label{fig:MR_MP1}
\end{figure}

In this appendix we show the mass-radius relations found using the APR EOS in figure \ref{fig:MR_APR} and the MP1 EOS in figure \ref{fig:MR_MP1}.

\begin{figure}
    \centering
\includegraphics[width=0.95\textwidth]{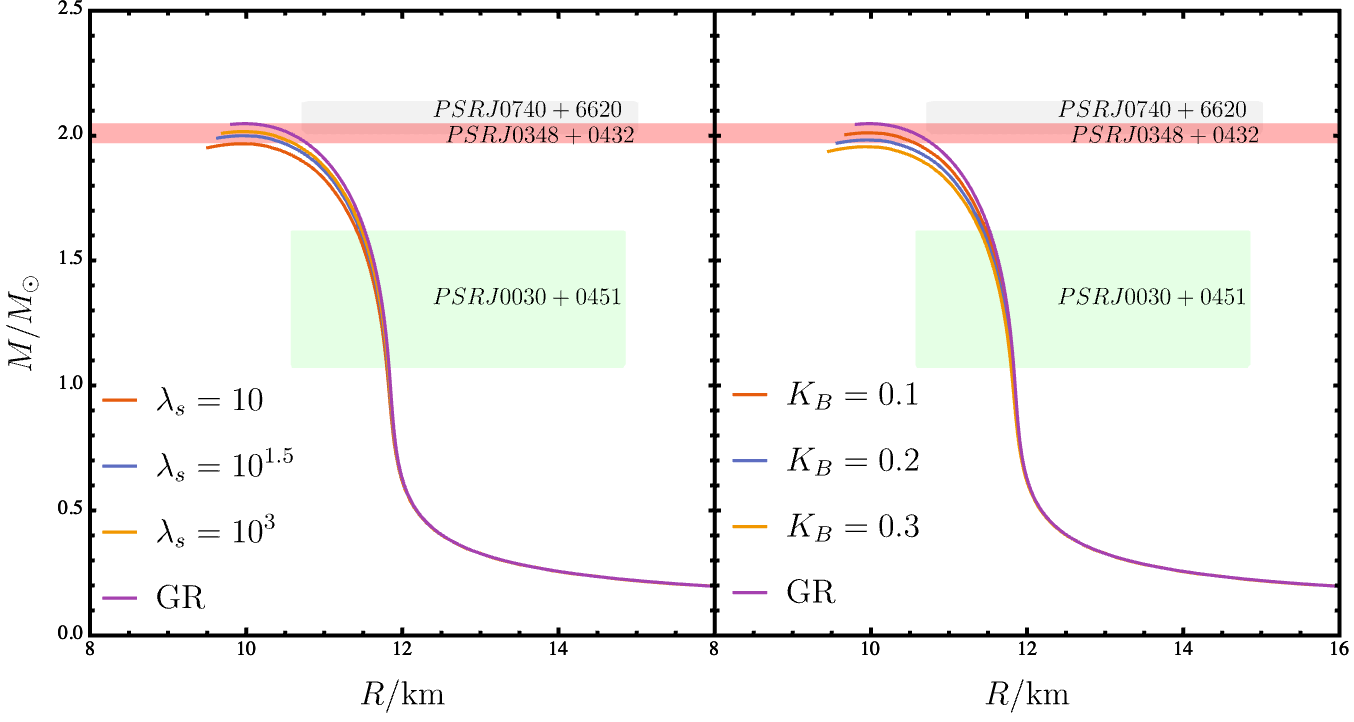}
    \caption{Same as figure~\ref{fig:MR_APR} but using the SLY4 EOS.}
    \label{fig:MR_SLY4}
\end{figure}

\clearpage
\bibliography{refs}
\end{document}